\documentclass[twocolumn,RNAAS]{aastex631}

\shorttitle{Stochastic Forces in REBOUNDx}
\shortauthors{Rein and Choksi}
\graphicspath{{./}{figures/}}

\begin{document}

\title{An Implementation of Stochastic Forces for the N-body code REBOUND}

\author[0000-0003-1927-731X]{Hanno Rein}
\affiliation{Department of Physical and Environmental Sciences, University of Toronto at Scarborough, Toronto, Ontario M1C 1A4, Canada}
\affiliation{Department of Astronomy and Astrophysics, University of Toronto, Toronto, Ontario, M5S 3H4, Canada}
\author[0000-0003-0690-1056]{Nick Choksi}
\affiliation{Astronomy Department, Theoretical Astrophysics Center, and Center for Integrative Planetary Science, University of California}

\begin{abstract}
We describe the implementation of a new module which can be used to simulate physical systems in which the motion of particles is affected by stochastic forces. 
Such forces are expected to be present in turbulent circumstellar disks or remnant planetesimal disks.
Our implementation offers a convenient way to generate correlated noise with a user-specified amplitude and auto-correlation time for each particle. 
The module has minimal memory requirements and is freely available within the REBOUNDx additional effects library. 
\end{abstract}

\section{Algorithm} \label{sec:algorithm}
Many problems in astrophysics are stochastic in nature \citep{Chandrasekhar1943}.
Examples from dynamics include the motion of stars in a cluster or galaxy and the orbital motion of planets embedded in gaseous protoplanetary disks or solid planetesimal disks \citep{Nelson2005, murray-clay_chiang_2006, oishi2007}.

Several authors have conducted $N$-body simulations of planets undergoing what is often referred to as stochastic migration \citep{AdamsBloch2009, ReinPapaloizou2009}.
In such long-term simulations, an additional stochastic force is applied to each particle to avoid having to run a more realistic but prohibitively slow
simulation generating the noise \text{ab initio}
in parallel to the N-body simulation.

In this research note, we present one implementation of such a stochastic force for the $N$-body code REBOUND \citep{ReinLiu2012} in the REBOUNDx additional effects library \citep{Tamayo2020}.
The algorithm is based on \cite{KASDIN} and similar to that used by \cite{ReinPapaloizou2009}.
It has several distinct advantages over other approaches.
\begin{enumerate}
    \item The perturbations are not implemented as instantaneous kicks but as a continuous force. This more closely resembles the actual force that particles feel in many physical systems. 
    \item The user can specify not only the amplitude of the stochastic forces, but also an auto-correlation time. 
    This is important if the auto-correlation time is similar to other timescales in the problem.
    For example, simulations of magnetohydrodynamic turbulence find  autocorrelation times that are comparable to the orbital period \citep{oishi2007, ReinPapaloizou2009}.
    \item The algorithm requires only one floating point number per coordinate as persistent memory in-between timesteps. This allows for a large number of particles experiencing stochastic forces, makes it easy to take snapshots of simulations, and aids reproducibility \citep{ReinTamayo2017}.
\end{enumerate}

\begin{figure*}[ht!]
\plotone{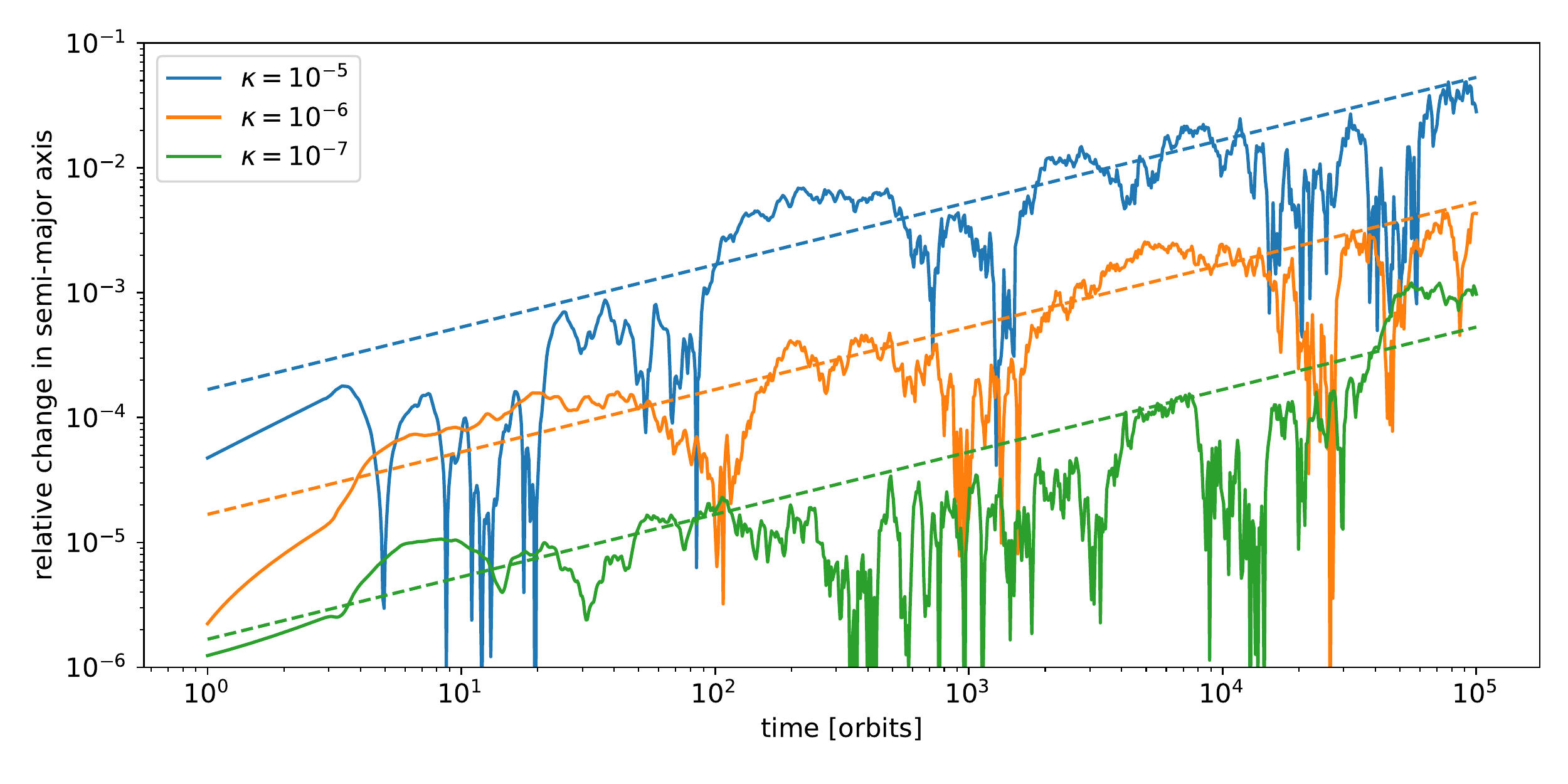}
\caption{The relative semi-major axis change of a planet undergoing stochastic migration with different diffusion parameters. The dashed lines are analytic predictions from \cite{ReinPapaloizou2009}.
\label{fig:plot}}
\end{figure*}

\section{Implementation}
We here describe the implementation of the \texttt{stochastic\_forces} module in REBOUNDx. 
This modules allows stochastic forces to be configured on a per-particle basis.

By default, it is assumed that particles are ordered by increasing semi-major axis, with the central object having a particle index of 0. 
Stochastic forces are then applied in both the radial and azimuthal directions (but not the vertical one). 
To turn on stochastic forces for a specific particle, one needs to set the standard deviation of the force $\kappa$ (\texttt{kappa}) to a finite value. 
The strength $\kappa$ is measured relative to the gravitational force from the central object.
For example, $\kappa = 10^{-6}$ (as expected for MHD turbulence; \citealt{oishi2007,ReinPapaloizou2009}) implies that the typical magnitude of the stochastic force is a million times weaker than the stellar gravitational force.
The auto-correlation function is modeled as an exponential with an e-folding timescale $\tau_\kappa$. We set $\tau_{\kappa}$ equal to the orbital period by default, but the user can specify an arbitrary time by setting the parameter \texttt{tau\_kappa}.

Whereas applying forces in the radial and azimuthal directions is useful for particles in orbits around other objects, we also provide the functionality to apply forces in Cartesian coordinates. 
To turn on stochastic forces in the $x$ direction for a particle, one needs to set a finite value to $\kappa_x$ (\texttt{kappa\_x}) as well as to $\tau_{\kappa,x}$ (\texttt{tau\_kappa\_x}). 
The parameters for the $y$ and $z$ directions are analogous.

There are two important considerations when choosing an integrator for systems in which stochastic forces are present. First, the timestep needs to be smaller than the auto-correlation time of the simulation to lead to physically meaningful results. Second, adaptive integrators such as IAS15 \citep{ReinSpiegel2015} do not work well because the integrator might choose prohibitively small timesteps. 
Integrators with a fixed timestep such as WHFast \citep{ReinTamayo2015} or Leap Frog are better suited. 
The stochastic forces described here can be used in combination with other forces in the REBOUNDx library (e.g. one can impose some stochasticity on top of smooth migration forces).

Our module uses the pseudo-random number generator within REBOUND.
To achieve reproducible results, one can set the random seed of the corresponding REBOUND simulation to a fixed value.

Our implementation and accompanying documentation is freely available in the REBOUNDx library at \url{https://github.com/dtamayo/reboundx}.

\section{Example}
In Figure~\ref{fig:plot} we plot the relative change of a planet subjected to stochastic forces in the radial and azimuthal directions.
We vary the amplitude of the stochastic forces $\kappa$ between $10^{-7}$ and $10^{-5}$. 
The numerical results match analytic predictions from \cite{ReinPapaloizou2009}, who provide a set of equations relating the strength of the perturbing forces to the time-averaged change in orbital parameters.

\begin{acknowledgments}
This work has been supported by the NSERC Discovery Grant RGPIN-2020-04513 and was made possible by the open-source projects 
\texttt{Jupyter} \citep{jupyter}, \texttt{iPython} \citep{ipython}, \texttt{matplotlib} and \citep{matplotlib, matplotlib2}.

\end{acknowledgments}

\bibliography{full}{}

\begin{thebibliography}{}
\expandafter\ifx\csname natexlab\endcsname\relax\def\natexlab#1{#1}\fi
\providecommand{\url}[1]{\href{#1}{#1}}
\providecommand{\dodoi}[1]{doi:~\href{http://doi.org/#1}{\nolinkurl{#1}}}
\providecommand{\doeprint}[1]{\href{http://ascl.net/#1}{\nolinkurl{http://ascl.net/#1}}}
\providecommand{\doarXiv}[1]{\href{https://arxiv.org/abs/#1}{\nolinkurl{https://arxiv.org/abs/#1}}}

\bibitem[{{Adams} \& {Bloch}(2009)}]{AdamsBloch2009}
{Adams}, F.~C., \& {Bloch}, A.~M. 2009, \apj, 701, 1381,
  \dodoi{10.1088/0004-637X/701/2/1381}

\bibitem[{{Chandrasekhar}(1943)}]{Chandrasekhar1943}
{Chandrasekhar}, S. 1943, Reviews of Modern Physics, 15, 1,
  \dodoi{10.1103/RevModPhys.15.1}

\bibitem[{Droettboom {et~al.}(2016)Droettboom, Hunter, Caswell, Firing,
  Nielsen, Elson, Root, Dale, Lee, Seppänen, McDougall, Straw, May, Varoquaux,
  Yu, Ma, Moad, Silvester, Gohlke, Würtz, Hisch, Ariza, Cimarron, Thomas,
  Evans, Ivanov, Whitaker, Hobson, mdehoon, \& Giuca}]{matplotlib2}
Droettboom, M., Hunter, J., Caswell, T.~A., {et~al.} 2016, matplotlib:
  matplotlib v1.5.1, \dodoi{10.5281/zenodo.44579}

\bibitem[{Hunter(2007)}]{matplotlib}
Hunter, J.~D. 2007, Computing In Science \& Engineering, 9, 90

\bibitem[{{Kasdin}(1995)}]{KASDIN}
{Kasdin}, N.~J. 1995, Proceedings of the IEEE, 83

\bibitem[{Kluyver {et~al.}(2016)Kluyver, Ragan-Kelley, P{\'e}rez, Granger,
  Bussonnier, Frederic, Kelley, Hamrick, Grout, Corlay, {et~al.}}]{jupyter}
Kluyver, T., Ragan-Kelley, B., P{\'e}rez, F., {et~al.} 2016, Positioning and
  Power in Academic Publishing: Players, Agents and Agendas, 87

\bibitem[{{Murray-Clay} \& {Chiang}(2006)}]{murray-clay_chiang_2006}
{Murray-Clay}, R.~A., \& {Chiang}, E.~I. 2006, \apj, 651, 1194,
  \dodoi{10.1086/507514}

\bibitem[{{Nelson}(2005)}]{Nelson2005}
{Nelson}, R.~P. 2005, \aap, 443, 1067, \dodoi{10.1051/0004-6361:20042605}

\bibitem[{{Oishi} {et~al.}(2007){Oishi}, {Mac Low}, \& {Menou}}]{oishi2007}
{Oishi}, J.~S., {Mac Low}, M.-M., \& {Menou}, K. 2007, \apj, 670, 805,
  \dodoi{10.1086/521781}

\bibitem[{P\'erez \& Granger(2007)}]{ipython}
P\'erez, F., \& Granger, B.~E. 2007, Computing in Science and Engineering, 9,
  21, \dodoi{10.1109/MCSE.2007.53}

\bibitem[{{Rein} \& {Liu}(2012)}]{ReinLiu2012}
{Rein}, H., \& {Liu}, S.-F. 2012, \aap, 537, A128,
  \dodoi{10.1051/0004-6361/201118085}

\bibitem[{{Rein} \& {Papaloizou}(2009)}]{ReinPapaloizou2009}
{Rein}, H., \& {Papaloizou}, J.~C.~B. 2009, \aap, 497, 595,
  \dodoi{10.1051/0004-6361/200811330}

\bibitem[{{Rein} \& {Spiegel}(2015)}]{ReinSpiegel2015}
{Rein}, H., \& {Spiegel}, D.~S. 2015, \mnras, 446, 1424,
  \dodoi{10.1093/mnras/stu2164}

\bibitem[{{Rein} \& {Tamayo}(2015)}]{ReinTamayo2015}
{Rein}, H., \& {Tamayo}, D. 2015, MNRAS, 452, 376

\bibitem[{{Rein} \& {Tamayo}(2017)}]{ReinTamayo2017}
---. 2017, MNRAS, 467, 2377

\bibitem[{Tamayo {et~al.}(2020)Tamayo, Rein, Shi, \& Hernandez}]{Tamayo2020}
Tamayo, D., Rein, H., Shi, P., \& Hernandez, D.~M. 2020, Monthly Notices of the
  Royal Astronomical Society, 491, 2885, \dodoi{10.1093/mnras/stz2870}

\end{thebibliography}
\bibliographystyle{aasjournal}

\end{document}